\documentclass[sigconf]{acmart}

\usepackage{soul}
\usepackage{framed}
\usepackage{graphicx}
\usepackage{float}
\usepackage{url}
\usepackage{rotating}

\usepackage{natbib}
\usepackage{geometry}
\usepackage{fleqn}

\usepackage{epsfig}

\usepackage{pgfplots, pgfplotstable}
\usepackage{scalefnt}
\usepackage{tikz}
\usetikzlibrary{shapes.gates.logic.US,trees,positioning,arrows}
\usetikzlibrary{shapes.multipart}
\usetikzlibrary{shapes,arrows,snakes}

\usepackage{alltt                                    
	, multirow
	,subfig
	, booktabs
	, listings
	, graphicx
	,float
	,cite
	,verbatim
	,mathtools
	,url
	,amsmath
}

\usepackage{verbatim}
\usetikzlibrary{arrows,shapes,backgrounds}
\usepgfplotslibrary{statistics}

\usetikzlibrary{shapes.gates.logic.US,trees,positioning,arrows}
\usetikzlibrary{shapes.multipart}
\usetikzlibrary{shapes,arrows}

\tikzstyle{vertex}=[ellipse,fill=black!25,minimum size=20pt, inner sep=0pt]
\tikzstyle{edge} = [draw,thin,-]
\tikzstyle{glabel} = [text width=1cm,text centered,font=\bf]
\pgfdeclarelayer{bg}    
\pgfsetlayers{bg,main} 

\usepackage{pgfplots, pgfplotstable}
\usepackage{tikz}
\pgfplotsset{compat=1.16}
\usepackage{pgfplotstable}
\usepackage{pgf-pie}

\newif\ifpienumberinlegend
\pgfkeys{/number in legend/.code=
    \expandafter\let\expandafter\ifpienumberinlegend
    \csname if#1\endcsname
    \ifpienumberinlegend

    \def\beforenumber##1\afternumber{}%
    \fi,
    /number in legend/.default=true
}

\makeatletter
\pgfplotsset{
    boxplot/hide outliers/.code={
        \def\pgfplotsplothandlerboxplot@outlier{}%
    }
}
\makeatother

\tikzstyle{vertex}=[ellipse,fill=black!25,minimum size=20pt, inner sep=0pt]
\tikzstyle{edge} = [draw,thin,-]
\tikzstyle{glabel} = [text width=1cm,text centered,font=\bf]
\pgfdeclarelayer{bg}    
\pgfsetlayers{bg,main} 

\usepackage{tcolorbox}

\usepackage{fancybox}
\usepackage{multirow}
\usepackage{flushend}
\usepackage{booktabs}
\usepackage{tabularx}
\usepackage{comment}
\usepackage{array}
\usepackage{mdframed}
\DeclareGraphicsExtensions{.pdf,.png}

\usepackage{subfig}

\usepackage{listings}
\usepackage{courier}
\usepackage{color}
\usepackage{hyphenat}

\usepackage{alltt                                    
	, multirow
	,subfig
	, booktabs
	, listings
	,float
	,verbatim
	,mathtools
	,url
	,amsmath
}
\usepackage{framed,lipsum}
\usepackage{pgfkeys}
\usepackage{graphics} 
\usepackage{graphicx}
\usepackage{pgfplots, pgfplotstable}
\pgfplotsset{compat=1.15}
\usepackage{tikz}
\usetikzlibrary{patterns}

\usepackage{algorithm2e}
\usepackage{algpseudocode}

\newcounter{o}
\usepackage{xspace}

\usepackage{tikz}
\usepackage{pgf-pie}
\usetikzlibrary{positioning,shadows}

\newcommand*\emptycirc[1][1ex]{\tikz\draw (0,0) circle (#1);} 
\newcommand*\halfcirc[1][1ex]{%
  \begin{tikzpicture}
  \draw[fill] (0,0)-- (90:#1) arc (90:270:#1) -- cycle ;
  \draw (0,0) circle (#1);
  \end{tikzpicture}}
\newcommand*\fullcirc[1][1ex]{\tikz\fill (0,0) circle (#1);} 

\definecolor{1c1}{RGB}{188,162,6}
\definecolor{1c2}{RGB}{137,129,80}
\definecolor{1c3}{RGB}{239,167,31}
\definecolor{1c4}{RGB}{88,194,241}
\definecolor{1c5}{RGB}{6,180,188}

\tikzset{mynode/.style={draw=white,solid,circle,fill=green,inner sep=1pt, thick,
text=black}}
\tikzset{arrow line/.style={dashed, line width= 2.5pt, color=#1}}

\def\bf{\textbf}

\usepackage{balance}

\normalsize

\usepackage{enumitem}

\usepackage{tikz}

\lstdefinestyle{inlinecode}{basicstyle={\ttfamily\scriptsize\bfseries}}

\newcommand{\urls}[1]{{\scriptsize\url{#1}}}

\usepackage{soul}
\usepackage[outercaption]{sidecap}    

\usepackage{enumitem}
\usepackage[T1]{fontenc}
\usepackage{pifont}

\definecolor{lightgray}{gray}{.92}

\usepackage{graphicx}
\usepackage{subfig}

\copyrightyear{2024} \acmYear{2024} \setcopyright{acmlicensed}
\acmConference[MSR '24]{21st International Conference on Mining Software Repositories}{April 15--16, 2024}{Lisbon, Portugal} 
\acmBooktitle{21st International Conference on Mining Software Repositories (MSR '24), April 15--16, 2024, Lisbon, Portugal} 
\acmDOI{10.1145/3643991.3645085} 
\acmISBN{979-8-4007-0587-8/24/04}

\begin{document}


\title{Enhancing User Interaction in ChatGPT: Characterizing and Consolidating Multiple Prompts for Issue Resolution}

\author{Saikat Mondal}
\affiliation{%
  \country{University of Saskatchewan, Canada}
    }
\email{saikat.mondal@usask.ca}

\author{Suborno Deb Bappon}
\affiliation{%
  \country{University of Saskatchewan, Canada}
    }
\email{suborno.deb@usask.ca}

\author{Chanchal K. Roy}
\affiliation{%
  \country{University of Saskatchewan, Canada}
  }
\email{chanchal.roy@usask.ca}

\begin {abstract}

Prompt design plays a crucial role in shaping the efficacy of ChatGPT, influencing the model's ability to extract contextually accurate responses. Thus, optimal prompt construction is essential for maximizing the utility and performance of ChatGPT. However, sub-optimal prompt design may necessitate iterative refinement, as imprecise or ambiguous instructions can lead to undesired responses from ChatGPT. Existing studies explore several prompt patterns and strategies to improve the relevance of responses generated by ChatGPT. However, the exploration of constraints that necessitate the submission of multiple prompts is still an unmet attempt.
In this study, our contributions are twofold.
First, we attempt to uncover gaps in prompt design that demand multiple iterations. In particular, we manually analyze $686$ prompts that were submitted to resolve issues related to Java and Python programming languages and identify \emph{eleven} prompt design gaps (e.g., missing specifications). Such gap exploration can enhance the efficacy of single prompts in ChatGPT.
Second, we attempt to reproduce the ChatGPT response by consolidating multiple prompts into a single one. We can completely consolidate prompts with \emph{four} gaps (e.g., missing context) and partially consolidate prompts with \emph{three} gaps (e.g., additional functionality).
Such an effort provides concrete evidence to users to design more optimal prompts mitigating these gaps.
Our study findings and evidence can -- (a) save users time, (b) reduce costs, and (c) increase user satisfaction.

\end {abstract}

\ccsdesc[400]{Large Language Model~Prompt design}
\ccsdesc[400]{Large Language Model~Prompt characterization}

\keywords{Prompt design, ChatGPT, prompt consolidation, qualitative analysis}




\maketitle

\section{Introduction}

\begin{figure}[!htb]
	\centering
	\includegraphics[width=2.7in]{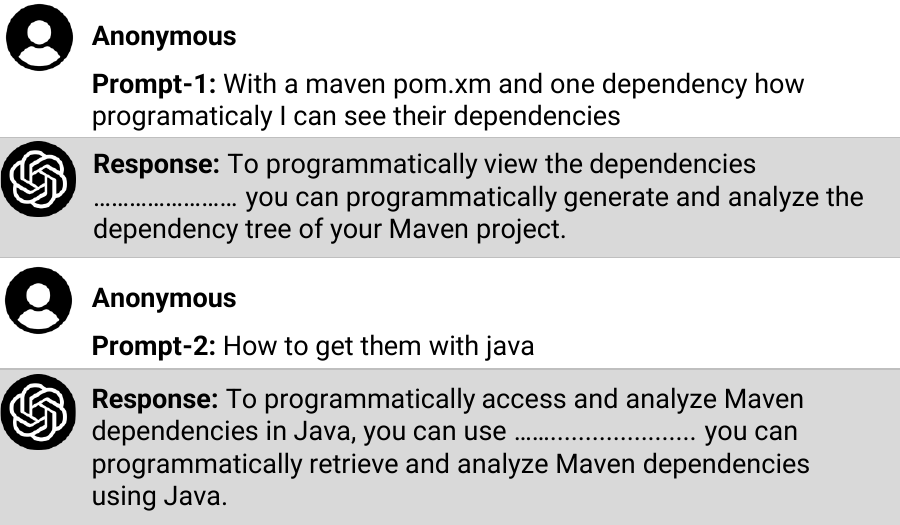}
	\caption{Example of a developer-ChatGPT conversation \citep{ChatGPTSharingURL1}.}
	\label{fig:prompt-example}
\end{figure}

Software developers are rapidly adopting large language models (LLMs) such as ChatGPT to generate code and other software engineering artifacts \citep{bommasani2021opportunities, bang2023multitask}.
A prompt is used to interact with ChatGPT to generate responses (e.g., Fig. \ref{fig:prompt-example}). However, the design of a prompt is crucial for getting appropriate responses from ChatGPT. It shapes the developer-ChatGPT conversations and serves as the foundation for obtaining accurate, relevant, and meaningful responses from ChatGPT. However, sub-optimal prompt design may necessitate iterative adjustment. A prompt with unclear and insufficient context can lead to undesired responses from ChatGPT. Therefore, it demands multiple prompts to obtain the desired responses from ChatGPT. Such a scenario -- (a) introduces complexity to user interactions, (b) costs time and money, and (c) poses the risk of undermining the model's efficiency, disrupting the user experience.

\begin{figure*}[htb]
	\centering
	\includegraphics[width=6.3in]{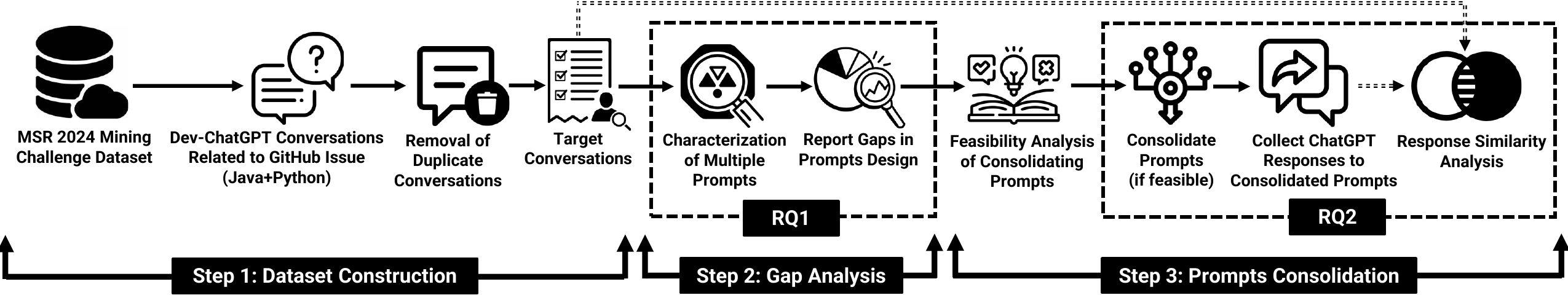}
	\caption{Schematic diagram of our conducted study.}
	\label{fig:schematic-diagram}
 \vspace{-2mm}
\end{figure*}

Several existing studies suggest prompt patterns designed to enhance human interaction with ChatGPT in the context of software engineering \citep{white2023prompt, white2023chatgpt, schmidtcataloging}. Users can improve the outputs of LLM conversations by employing such patterns. Elnashar et al.~\citep{elnasharprompt} experiment with various prompting strategies to generate Python code with ChatGPT. Their study suggests that prompting multiple times for code and selecting the best code among them is comparable to top-rated human solutions. However, understanding the gaps that drive iterative prompt refinement is crucial for optimizing the model's performance, streamlining user interactions, and ensuring more effective problem resolution.

In this study, we conduct a manual investigation to characterize multiple prompts submitted to resolve GitHub issues related to Java and Python programming languages. We attempt to explore constraints that necessitate the submission of multiple prompts. Additionally, we performed a feasibility study to identify prompts that could be consolidated into a single prompt, reproducing the original ChatGPT response.
In particular, we answer two research questions in this study as follows.

\smallskip
\noindent\textbf{RQ1.} \textbf{Can we identify the gaps in prompt design requiring multiple iterations to enhance the effectiveness of single prompts in ChatGPT?}
We manually analyze $85$ conversations that contain $686$ prompts and identify \emph{eleven} gaps in prompts, where \emph{missing specifications} and \emph{additional functionality} are seen as the most frequent gaps in the prompt design.

\noindent\textbf{RQ2.} \textbf{To what extent can prompt consolidation reproduce chatGPT responses?}
We can completely consolidate the prompts and reproduce original responses by minimizing four gaps (Missing Specifications, Wordy Response, Missing Context, and Miscellaneous) and partially by minimizing three gaps (Incremental Problem Solving, Exploring Alternative Approaches, and Additional Functionality). The remaining four gaps (e.g., Inaccurate/Untrustworthy Response) could not be mitigated since they depend on the prompt response.

\smallskip
\noindent\textbf{Replication Package} that contains our manually identified gaps in prompt design and consolidated prompts can be found in our online appendix \citep{replicationPackage}.

\section{Study Methodology}
\label{methodilogy}
In Fig. \ref{fig:schematic-diagram}, we describe our overall methodology to answer the two research questions. We describe the steps below.

\smallskip
\textbf{Step 1: Data Collection and Preprocessing.}
Table \ref{table:dataset-summary} shows the summary of our collected dataset. We downloaded the MSR challenge 2024 dataset \citep{xiao2023devgpt}, which encompasses 17,913 prompts, including source code, commits, issues, pull requests, discussions, and Hacker News threads. In particular, we collect developer-ChatGPT conversations related to GitHub issues in two popular programming languages -- Java and Python from all nine snapshots. We get a total of 763 conversations (225 Java + 538 Python). Our initial screening reveals that snapshots contain duplicate conversations. We then filter out the duplicate conversations using \texttt{ChatgptSharing-URL}. Finally, we get 108 distinct conversations (27 Java + 81 Python), where
85 conversations have multiple prompts.

\begin{table}[htb]
	\centering
    \caption{Summary of the study dataset}
    \label{table:dataset-summary}
    \resizebox{3.2in}{!}{%
    \begin{tabular}{l|c|c|c} \hline
            \textbf{Language} & \textbf{\# of Conversations} & \textbf{\begin{tabular}[c]{@{}l@{}}\# of Unique \\ Conversations\end{tabular}} & \textbf{\begin{tabular}[c]{@{}l@{}}Multiple-Prompts\\ Conversations\end{tabular}} \\ \hline
            \textbf{Python}   &   538   &   81     &  67        \\ \hline
            \textbf{Java}     &  225    &   27     &   18       \\ \hline
            \textbf{Total}     &  763   &   108   &   85       \\ \hline

    \end{tabular}
   }
 \end{table}

\smallskip
\textbf{Step 2: Gap Analysis in Prompt Design.}
We get $85$ conversations from Step 1 (Fig. \ref{fig:schematic-diagram}) that contain $686$ prompts. We, the first two authors of the paper, conduct a careful manual investigation of these conversations to find the gaps in prompt design. In particular, we attempt to characterize why multiple prompts were required for GitHub issue resolution. We randomly select 20 conversations (10 Java + 10 Python) from our selected dataset and analyze \& label them together. We meticulously analyze the prompts and responses for a given conversation to explore why users submit multiple prompts. Such analysis enables us to come up with a common understanding. We then analyze the prompts in the remaining conversations and label them individually. Finally, we resolved a few disagreements through discussion. 
Please note that prompts in a single conversation can have multiple labels.

\smallskip
\textbf{Step 3: Prompts Consolidation.}
In this step, we first assess the feasibility of consolidating ChatGPT prompts. Feasible prompts are then consolidated into a single prompt. 
We manually consolidate the prompts utilizing the ChatGPT-3.5\footnote{https://chat.openai.com} browser interface. We then proceed with ChatGPT-3.5 to reproduce the original response using these consolidated prompts. We also randomly select 50\% of consolidated prompts and attempt to reproduce the original response using ChatGPT-4  to examine the generalizability of the results.
We then examine whether the original prompt's response and response from the consolidated prompt are equivalent or not. The success in reproducing ChatGPT's original response validates that users can achieve desired outcomes with more optimal prompts.

\begin{table*}[htb]
	\centering
    \caption{Summary of Characterization of Multiple Prompts and Gap Analysis in Prompt Design}
    \label{table:prompt-design-gaps-summary}
    \resizebox{7in}{!}{%
    \begin{tabular}{p{8cm}|p{11cm}|c} \hline 
    
\textbf{Gaps/Reasons with Description} & \textbf{Examples (as submitted by developers)} & \textbf{Count} \\ \hline

\textbf{(1) Missing Specifications.} 
Prompts miss explicit information (e.g., programming language, operating system, test cases) and requirements (e.g., asking code).   
& 
\emph{\underline{Prompt 1}}: With a maven pom.xm and one dependency, how
programmatically can I see their dependencies? \emph{\underline{Prompt 2}}: How to get them with \textbf{java}. [see Fig. \ref{fig:prompt-example}] \citep{ChatGPTSharingURL1}
&   34  \\ \hline

\textbf{(2) Different Use Cases.} 
Different prompts ask for responses to resolve different problems in a single conversation.   
& \emph{\underline{Prompt 1}}: Using the Python ast module, \textbf{how can I access the docstring} for a function? \emph{\underline{Prompt 2}}:   In Python, \textbf{how can I turn a multiline string into a triple-quoted string} literal easily? \citep{ChatGPTSharingURL2} 
& 25 \\ \hline

\textbf{(3) Incremental Problem Solving.}   
Users break down a problem into several components and submit prompts sequentially to seek solutions for each part.   
& \emph{\underline{Prompt 2}}: Use numpy to run all of the vector cosine calculations in one go rather than one at a time.
\emph{\underline{Prompt 3}}: \textbf{Now run GitHubenchmark again} with 10,000 vectors. \citep{ChatGPTSharingURL34}
& 10 \\ \hline

\textbf{(4) Exploring Alternative Approaches.}   
Users ask for alternative solutions when they are not satisfied with a response.   
&  \emph{\underline{Prompt 4}}: Can you use anything in scipy or pandas or similar to beat these results?
\emph{\underline{Prompt 5}}: \textbf{Try some other distance calculations}, I want to find the fastest possible way to compare these vectors. \citep{ChatGPTSharingURL34}
& 5 \\ \hline

\textbf{(5) Wordy Response.}   
Users submit prompts to get an optimal response when the received response is wordy.   
&  \emph{\underline{Prompt 1}}: Figure out how to solve this GitHub issue: \href{https://github.com/AntonOsika/gpt-engineer/issues/294}{[GitHub-Issue-Link]} by reviewing the code at this repo: \href{https://github.com/AntonOsika/gpt-engineer}{[Repo-Link]}.
\emph{\underline{Prompt 2}}: \textbf{Summarize what this potential resolution does} from a high level. \citep{ChatGPTSharingURL5}
& 2 \\ \hline

\textbf{(6) Additional Functionality.}   
Prompts ask for additional functionality (e.g., features) in addition to the existing solution.   
& \emph{\underline{Prompt 1}}: with flask in Python and rabbit .......... and otherwise response with a timeout error
\emph{\underline{Prompt 2}}: \textbf{can you give an example of the flask side of things} where we receive a request and wait for a message on rabbitMQ before we send the response \citep{ChatGPTSharingURL6}
& 33 \\ \hline

\textbf{(7) Erroneous Response.}   
Prompts are submitted to seek an accurate solution when the received response is identified as error-prone.   
& \emph{\underline{Prompt 2}}: now create a simple GET endpoint............ with domain \href{https://bffd-174-64-129-70.ngrok-free.app}{[Link]}.
\emph{\underline{Prompt 3}}: 
\{
``\textbf{error}'':``AutoReconnect error: ac-v4z8kcr-lb.fbpeqkx.mongodb.net:27017: [SSL: CERTIFICATE\_VERIFY\_FAILED] certificate verify failed: certificate has expired (\_ssl.c:997''
\} \citep{ChatGPTSharingURL7}
& 21 \\ \hline

\textbf{(8) Missing Context.}   
Prompts lack sufficient background information or details for the ChatGPT model to generate a contextually appropriate response.   
& \emph{\underline{Prompt 1}}: I have 2 different versions of a sqlite database......................... Can you show me how I can do this?
\emph{\underline{Prompt 2}}: \textbf{This is the table scheme} of favorites: CREATE TABLE favorites (name, URL, mode, image, duration, quality) \citep{ChatGPTSharingURL8}
& 12 \\ \hline

\textbf{(9) Clarity of Generated Response.}   
Users submit prompts when they encounter difficulty in comprehending any aspect of the received response.
& \emph{\underline{Prompt 2}}: Now run that against ......... List[Tuple[AST, Optional[str]]]:
\emph{\underline{Prompt 3}}: \textbf{Figure out why it's not working correctly}, and write and demonstrate snippets of code to fix that \citep{ChatGPTSharingURL9}
& 26 \\ \hline

\textbf{(10) Inaccurate/Untrustworthy Response.}   
Responses fail to operate as intended, cannot resolve the problem, or the generated output lacks reliability for further consideration.   
& \emph{\underline{Prompt 3}}: I think ............ this condition instead?
\emph{\underline{Prompt 4}}: the Python documentation does not contain any references to os.timerfd\_create though. \textbf{Are you sure this really exists?} \citep{ChatGPTSharingURL10}
& 6 \\ \hline

\textbf{(11) Miscellaneous.} 
Users submit prompts containing expressions of gratitude (e.g., thank you) in response to ChatGPT responses and null queries (unintentionally or out of curiosity).   
& \emph{\underline{Prompt 1}}: Write me a function .......... the translation matrix instead of multiplication. 
\emph{\underline{Prompt 2}}: \textbf{Thank you} \citep{ChatGPTSharingURL11}
& 2 \\ \hline
   
    \end{tabular}
   }
 \end{table*}

  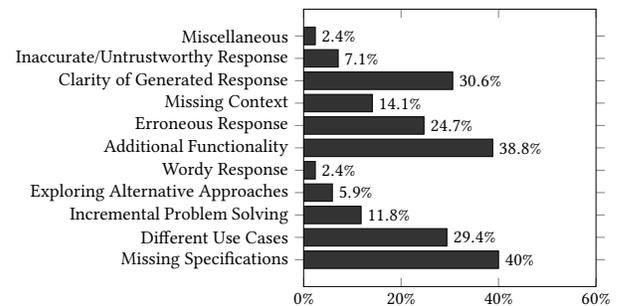
\begin{figure}[!htb]
\centering
        \resizebox{3.2in}{!}{%
    	\begin{tikzpicture}
        	\begin{axis}
            [
            height = 2.5in,
            width = 2.6in,
            xmin=0, xmax=60,
            xbar,
            ytick = data,
            enlarge y limits=0.12,
        	enlarge x limits=false,
	        nodes near coords,
            xtick={0,20,...,60},
            xticklabels={0\%,20\%,40\%,60\%},
            bar width=0.3cm,
            symbolic y coords = {
            Missing Specifications,
            Different Use Cases,
            Incremental Problem Solving,
            Exploring Alternative Approaches,
            Wordy Response,
            Additional Functionality,
            Erroneous Response,
            Missing Context,
            Clarity of Generated Response,
            Inaccurate/Untrustworthy Response,
            Miscellaneous
            },
            legend style={at={(0.5,-0.20)},
        	font=\footnotesize,
            anchor=north,legend columns=-1},
            nodes near coords style={rotate=0,  anchor=west}, 
        	nodes near coords =\pgfmathprintnumber{\pgfplotspointmeta}\%
            ]
            \addplot[xbar,fill=black!80] coordinates {
            (40,Missing Specifications)
            (29.4,Different Use Cases)
            (11.8,Incremental Problem Solving)
            (5.9,Exploring Alternative Approaches) 
            (2.4,Wordy Response) 
            (38.8,Additional Functionality) 
            (24.7,Erroneous Response) 
            (14.1,Missing Context)
            (30.6,Clarity of Generated Response)
            (7.1,Inaccurate/Untrustworthy Response)
            (2.4,Miscellaneous)
            };
           \end{axis}
        \end{tikzpicture}
        }
\caption{The percentage of each prompt design gap.}
\vspace{-5mm}
\label{fig:rank-gaps-prompt-design}
\end{figure}

\section{Study Findings}
\label{sec:study-findings}
We ask two research questions in this study. In this section, we answer them carefully with the help of our empirical and qualitative findings as follows:

\subsection{Answering RQ1: Characterization of Multiple Prompts and Gap Analysis in Prompt Design}
\label{subsec:gaps-analysis}

In this section, we present our findings on the gap analysis found in prompt design that necessitates multiple prompts in a conversation.

Table \ref{table:prompt-design-gaps-summary} summarizes why multiple prompts were submitted in a single conversation and reports the gaps in prompt design with examples. Our manual analysis identified a total of \emph{eleven} gaps in the prompt design.
Fig. \ref{fig:rank-gaps-prompt-design} shows the ratio of each prompt design gap in our selected dataset. We see that \emph{missing specifications} (i.e., 40\%) and \emph{additional functionality} (i.e., 38.8\%) requirements are the most frequently observed gaps in prompt design, necessitating the submission of multiple prompts.

\begin{table}[htb]
	\centering
    \caption{Feasibility of Prompt Consolidation (\fullcirc[0.7ex] $\rightarrow$ Complete, \halfcirc[0.7ex] $\rightarrow$ Partial \& \emptycirc[0.7ex] $\rightarrow$ No Consolidation)}
    \label{table:consolidation-feasibility}
    \resizebox{2.2in}{!}{%
    \begin{tabular}{lc} \toprule
    
        \textbf{Gap/Reason} & \textbf{Consolidation} \\ \hline

        Missing Specifications &  \fullcirc\\
        Different Use Cases &  \emptycirc\\
        Incremental Problem Solving &  \halfcirc\\
        Exploring Alternative Approaches &  \halfcirc\\ 
        Wordy Response &  \fullcirc\\
        Additional Functionality & \halfcirc\\
        Erroneous Response &  \emptycirc\\
        Missing Context &  \fullcirc\\
        Clarity of Generated Response &  \emptycirc \\
        Inaccurate/Untrustworthy Response &  \emptycirc\\
        Miscellaneous &  \fullcirc\\ \bottomrule

    \end{tabular}
   }
   \vspace{-4mm}
 \end{table}

\subsection{Answering RQ2 : Prompt Consolidation}
\label{sebsec:prompt-consolidation}

Our qualitative analysis identifies several gaps in prompt design (Section \ref{subsec:gaps-analysis}). This section attempts to mitigate these gaps and consolidate multiple prompts into one. Table \ref{table:consolidation-feasibility} summarizes the consolidation feasibility of each of the identified gaps. According to our analysis -- (a) four gaps could be completely mitigated (\fullcirc[0.7ex]) by consolidating prompts, (b) three can be partially mitigated (\halfcirc[0.7ex]), and (b) four could not be mitigated (\emptycirc[0.7ex]). 
Table \ref{table:consolidation-examples} shows a few representative consolidation examples (the complete list can be found in our online appendix \citep{replicationPackage}). 
In particular, we can consolidate 127 prompts into 52 and reproduce the original responses using ChatGPT-3.5. Then, half of the consolidated prompts (26 out of 52) were randomly selected to reproduce the original responses using ChatGPT-4. ChatGPT-4 can reproduce the original responses perfectly (responses can be found in our online appendix \citep{replicationPackage}). Such findings confirm the generalizability of the results.
However, four gaps were unsuccessfully mitigated because they (a) depend on the ChatGPT response or (b) discuss different use cases. For example, users cannot estimate the accuracy of the response before analyzing it.
Therefore, the ``Inaccurate/Untrustworthy Response'' gap could not be minimized. However, prompts with gaps like ``Missing Specifications'' and ``Additional Functionalities'' could be consolidated completely/partially. Such evidence can guide users in designing optimal prompts.

\section{Key Findings \& Guidelines} 
\label{sec:key-findings}

Our study provides several insights into why multiple prompts are submitted in one conversation. These insights may guide the users to improve the prompt design and increase the chance of getting appropriate responses with a minimal number of prompts.

    \textbf{(F1) Missing specifications hurt.} 
    Our manual analysis reveals that the most frequent gap in prompt design that demands multiple prompts is missing required specifications (e.g., programming language, sample input-output). Therefore, users should be cautious about adding the required specifications to get target responses in the lowest number of prompts.

    \textbf{(F2) Prompt language matters.} 
    Prompts should be submitted in English (if possible), and combining multiple languages in a single prompt is discouraged. ChatGPT primarily performs best with English prompts, and responses may be less accurate or appropriate when prompts are submitted in languages other than English. According to our analysis, conversations typically consist of 17.8 prompts on average (average token count = 5755.2) when the language is non-English or multi-language. In contrast, the corresponding statistic is 6.9 prompts (average token count = 1567.2) when the language of the prompts is English.

    \textbf{(F3) Prevent wordy response.} 
    Consider specifying the response length (e.g., number of words or sentences) to prevent wordy responses. Additionally, using a few key terms such as concise, summarized, short, or simpler may be helpful.

    \textbf{(F4) Generate robust code specifying the potential weakness.} 
    In the prompt, when asking for code, you can warn ChatGPT about potential coding errors or exceptions that may appear (if you are aware of them). Such a strategy helps the model generate robust code that can handle those errors/exceptions. 


\section{Related Work}
\label{sec:literature-review}

Several studies have attempted to improve prompt design with the aim of enhancing ChatGPT responses in the context of software engineering \citep{white2023prompt, white2023chatgpt, schmidtcataloging}.

White et al.~\citep{white2023prompt} introduce a prompt pattern catalog to enhance prompt engineering techniques and solve common problems when conversing with LLMs. This study also presents a framework for documenting patterns to structure prompts for diverse domains and insights into building prompts by combining multiple patterns. In another study, White et al.~\citep{white2023chatgpt} suggest prompt design techniques as patterns for software engineering tasks. They focus on automating common software engineering activities (e.g., generating API specifications from requirements). 

\begin{table}[htb]
	\centering
    \caption{Prompt Consolidation Examples}
    \label{table:consolidation-examples}
    \resizebox{3.2in}{!}{%
    \begin{tabular}{p{10cm}} \hline

    \emph{\underline{Prompt 1}}: With a maven pom.xm and one dependency how programmatically I can see their dependencies 
    
    \emph{\underline{Prompt 2}}: How to get them with java  \\

    \emph{\underline{\textbf{Consolidated Prompt}}}: How can I programmatically view the dependencies of a Maven project specified in the pom.xml file using Java?  \\ \hline 

    \emph{\underline{Prompt 1}}: I have 2 different versions of a SQLite database. The names are `favorites old.db' and `favorites.db'.
    I want to merge the content of the table favorites from the file `favorites old.db' into `favorites.db'. Skipping rows that are already in there.
    I am using DB Browser for SQLite, but if it is not possible with that, I have also Python I can use.Can you show me how I can do this?
    
    \emph{\underline{Prompt 2}}: This is the table scheme of favorites: CREATE TABLE favorites (name, URL, mode, image, duration, quality) \\
    
    \emph{\underline{\textbf{Consolidated Prompt}}}: Merge the `favorites' table from `favorites old.db' into `favorites.db' in SQLite while skipping duplicate rows. Provide instructions using either DB Browser for SQLite or Python. The table scheme for 'favorites' is:
    CREATE TABLE favorites (name, url, mode, image, duration, quality). \\ \hline

    \end{tabular}
   }
   \vspace{-4mm}
 \end{table}

Schmidt et al.~\citep{schmidtcataloging} address the disruptive impact of LLMs, such as ChatGPT, across various domains (e.g., education, medicine, and software engineering). They emphasize the importance of codifying prompt patterns for prompt engineering, offering a more disciplined and repeatable approach to interacting with and evaluating LLMs. It provides examples of prompt patterns that enhance human interaction with LLMs, specifically in software engineering. 

Elnashar et al.~\citep{elnasharprompt} experiment with various prompting strategies to generate Python code utilizing ChatGPT. The results indicate that choosing from the top solutions generated by ChatGPT is comparable to the best human solution on Stack Overflow. The study suggests that prompting multiple times for code and selecting the best solution from the generated options is a promising approach to assist human software engineers in finding optimal solutions.

The studies discussed above attempt to support prompt design to get a better response from LLMs such as ChatGPT. However, to the best of our knowledge, (a) characterizing multiple prompts to identify gaps in prompt design that necessitate multiple prompts and (b) analyzing the feasibility of consolidating multiple prompts into a single one that can replicate the response were not investigated before, making our study novel.

\section{Conclusion \& Future Work}
\label{sec:conclusion}

The effectiveness of ChatGPT significantly relies on prompt design, impacting its ability to produce contextually accurate responses. 
Optimal prompt construction is crucial for maximizing utility and performance. Suboptimal designs may require iterative refinement, as imprecise instructions can lead to undesired outputs. In this study, we first manually analyze $85$ conversations that contain $686$ prompts to identify the gaps that necessitate multiple prompts. Our careful analysis identifies eleven gaps where missing specifications and additional functionality are the top contributing gaps. Then, we analyze the feasibility of consolidating prompts that could reproduce the ChatGPT response. According to our analysis, four gaps could be completely, and three could be partially minimized. The insights of such findings can assist users in designing better prompts. We also offer evidence-based guidelines to design optimal prompts to save time, reduce costs, and increase user satisfaction.

In the future, we plan to introduce tool supports to identify gaps in prompt design and offer guidance to mitigate gaps to support the desired response with a minimum number of prompts.

\smallskip
\noindent\textbf{Acknowledgment:}
This research is supported in part by the Natural Sciences and Engineering Research Council of Canada (NSERC), and by the industry-stream NSERC CREATE in Software Analytics Research (SOAR).



\balance
\bibliographystyle{ACM-Reference-Format}
\bibliography{reference}

\end{document}